\documentclass[prl,twocolumn,showpacs,preprintnumbers,superscriptaddress,amsmath,amssymb]{revtex4}
\usepackage{graphicx}
\usepackage{subfigure}
\usepackage{mathrsfs}
\usepackage{amsfonts}
\usepackage{amsmath}
\usepackage{color}
\usepackage[colorlinks,linkcolor=blue,citecolor=blue]{hyperref}
\begin{document}

\title{Stochastic Thermodynamics of a Particle in a Box}
\author{Zongping Gong}
\affiliation{School of Physics, Peking University, Beijing 100871, China}
\affiliation{Department of Physics, University of Tokyo, 7-3-1 Hongo, Bunkyo-ku, Tokyo 113-0033, Japan}
\author{Yueheng Lan}
\email{lanyh@mail.tsinghua.edu.cn}
\affiliation{Department of Physics, Tsinghua University, Beijing 100084, China}
\affiliation{Collaborative Innovation Center of Quantum Matter, Beijing 100871, China}
\author{H. T. Quan}
\email{htquan@pku.edu.cn}
\affiliation{School of Physics, Peking University, Beijing 100871, China}
\affiliation{Collaborative Innovation Center of Quantum Matter, Beijing 100871, China}
\date{\today}

\begin{abstract}
The piston system (particles in a box) is the simplest and paradigmatic model in traditional thermodynamics. However, the recently established framework of stochastic thermodynamics (ST) fails to apply to this model system due to the embedded singularity in the potential. In this Letter we study the stochastic thermodynamics of a particle in a box by adopting a novel coordinate transformation technique. Through comparing with the exact solution of a breathing harmonic oscillator, we obtain analytical results of work distribution for an arbitrary protocol in the linear response regime, and verify various predictions of the Fluctuation-Dissipation Relation. %, which are shown to be in good agreements with the numerical results.
When applying to the Brownian Szilard's engine model, we obtain the optimal protocol $\lambda_t = \lambda_0 2^{t/\tau}$ for a given sufficiently long total time $\tau$. Our study not only establishes a paradigm for studying ST of a particle in a box, but also bridges the long-standing gap in the development of ST.
%We study the Brownian Szilard engine (BSE), a paradigmatic model of information thermodynamics governed by Langevianian dynamics. By making proper transformation of variables, the BSE model is mapped to a collision-free and boundless diffusive system, which enables us to obtain analytical results for a class of systems with rigid-wall potential. A reduced Feynman-Kac equation (RFKE) only in terms of momentum is derived under the frequent collision approximation, which turns out to be quite similar to the overdamped FKE of a breathing Brownian harmonic oscillator. Some analytical results in the linear response regime are obtained, and are shown to be in good agreements with the numerical results of stochastic simulation. Our study establish a new paradigm for studying stochastic thermodynamics with a time-dependent rigid wall potential.
\end{abstract}
\maketitle

\emph{Introduction.---} When opening any textbook of thermodynamics \cite{Callen_1985}, the piston system \cite{Piston}, or the classical ideal gas inside a rigid-wall potential is the simplest and an archetypal model used to illustrate various thermodynamic processes and cycles. %, for example the Carnot cycle.
In the context of traditional thermodynamics, due to the macroscopic size of the system, fluctuations are usually vanishingly small. % and averages of thermodynamic variables are sufficient to describe the system.
There work and heat are phenomenological variables and the microscopic equation of motion (EOM) is not directly relevant.
%Because of the lack of knowledge about the definition of work on trajectory level, one cannot study the fluctuations and its related fluctuation theorems, and beyond the linear response regime, no exact results were known \cite{Jarzynski_2011}. Accordingly, usually only properties of the quasistatic processes and the close-to-equilibrium (linear response regime) processes are considered.

When considering a small system, however, fluctuations become important and the EOM becomes essential \cite{Jarzynski_2011}. In recent years, substantial developments in the field of nonequilibrium thermodynamics in small systems \cite{Bustamante_2005} have been made. One of them is the formulation of the so-called stochastic thermodynamics (ST) \cite{Sekimoto_2010,Seifert_2012,Esposito_2015}, where stochastic dynamics is incorporated into thermodynamics. For small systems, e.g., a Brownian particle in a controllable potential, a coherent framework of thermodynamics at the trajectory level is constructed. Fluctuating thermodynamic variables, such as work, heat and entropy production, are identified as functionals of individual trajectories \cite{Uhlenbeck_1963,Hunter_1993,Jarzynski_1997,Sekimoto_1998}, based on which one can in principle calculate their distributions in arbitrary driven processes \cite{Speck_2004,Imparato_2005}, and thus go beyond the traditional thermodynamics. %When doing average over these trajectories, one can reproduce the known results in traditional thermodynamics, e.g., the minimum work principle $\left\langle W \right\rangle \ge \Delta F$.
In the linear response regime, the work distribution is Gaussian and satisfies the Fluctuation-Dissipation relations (FDRs) \cite{Hendrix_2001,Speck_2004}. What is more, even in arbitrarily far from equilibrium processes, some exact fluctuation relations concerning work, heat and entropy production %about the fluctuating thermodynamic variables
are discovered \cite{Harris_2007,Jarzynski_1997,Crooks_1998,Seifert_2005,BK_1981a,BK_1981b,Jarzynski_2007}. %such as Jarzynski equality \cite{Jarzynski_1997}, Crooks Fluctuation Theorems \cite{Crooks_1998}, and Fluctuation Theorem of total entropy production \cite{Seifert_2005} (there are some other relevant relations \cite{BK_1981a,BK_1981b,Jarzynski_2007}).
Experimentally, these fluctuation relations have been verified in various systems including a Brownian particle in a soft-wall potential \cite{Blickle_2006,Imparato_2007,Pak_2015,Ciliberto_2013}, exemplified by a charged colloidal particle trapped by an optical tweezer. The essential point of these developments in thermodynamics is the microscopic definition of work, heat and entropy at the trajectory level.

However, %it is easy to notice that the widely accepted
the usual microscopic definition of work $W[x_t] = \int dt \partial_t V_t(x_t)$ \cite{Jarzynski_1997,Sekimoto_1998} (see \cite{Sekimoto_2010,Jarzynski_2007,Rubi_2008a,Peliti_2008b,Peliti_2008a,Rubi_2008b,Jordan_2008,Rubi_2008c,Silbey_2009} for discussions and debates) is not applicable to the piston system, due to singularities in the rigid-wall potential, where work is done during discrete collisions of the particle with the moving piston \cite{Lua_2005,Bena_2005}. Previously, there are studies about work distributions of piston systems in nonequilibrium processes, but either with no contact with a heat bath \cite{Lua_2005,Bena_2005,Quan_2012}, or with no relevance to Brownian dynamics \cite{Hatano_1998,Baule_2006,Engel_2013,Broeck_2015}. The solution to the piston system becomes a ``missing puzzle piece" in ST. %As \textcolor{red}{is} mentioned, the piston system is the archetypal model of thermodynamics, and is of special importance. But unfortunately, the newly developed  \textcolor{red}{various techniques of the ST} cannot be directly applied to the piston system.
Possibly for lack of efficient ways of studying ST in a piston system, finite-time thermodynamics of the famous Brownian Szilard's engine (BSE) \cite{Hatano_1998,Broeck_2015,Szilard_1929,Leff_2003,Maruyama_2009,Dunkel_2014} remains unexplored so far. Hence, how to extend the framework of ST to the piston system becomes one of the most challenging problems in this field.

In this Letter, we try to %bridge the long-standing gap in the development of \textcolor{red}{ST by extending}
extend the previous framework of ST to include the %expanding
rigid-wall potential. We introduce a novel approach of coordinate transformation to study the ST in an isothermal piston. In this way, under certain conditions, the isothermal piston model is found to be highly similar to an isothermal breathing harmonic oscillator (HO) \cite{Speck_2011,Kwon_2013}, one of the very few models whose work distribution in an arbitrary process can be calculated analytically \cite{Kwon_2013}. Since exactly solvable models play an important role in statistical mechanics, considering the special role and the ubiquity of piston systems in thermodynamics, we believe that our work not only significantly extends the applicability of ST, but also has pedagogical value. We also note that the rigid-wall potential is accessible in current experiments \cite{Raizen_2005,Gaunt_2013}, so our findings could possibly be tested.

\emph{Model setup.---}Consider a single Brownian particle confined in a one-dimensional piston with its left boundary fixed while the right one movable. The mass of the particle is denoted by $m$, the left and right boundaries are at the origin $x = 0$ and $x = \lambda_t$ ($0 \leq t \leq \tau$) respectively. The piston system is coupled to a heat bath with inverse temperature $\beta$, so the motion of the Brownian particle can be described by the following underdamped Kramers-Langevin equation \cite{Sekimoto_2010}
\begin{equation}
\dot x = \frac{p}{m} \;\;, \;\; \dot p = -\gamma \frac{p}{m} + \sqrt{\frac{2 \gamma}{\beta}} \eta_t + I_c,
\label{Langevin}
\end{equation}
where $(x,p) \equiv \Gamma$ is the particle's position-momentum coordinate in the phase space, $\gamma$ is the viscous friction coefficient that characterizes the coupling strength between the piston system and the heat bath, $\eta_t$ is the standard Weiner process satisfying $\langle \eta_t \eta_{t'} \rangle = \delta (t - t')$ and $\eta_t dt \thicksim N(0,dt)$ (normal distribution with mean zero and variance $dt$), and $I_c$ is the collision term responsible for the collisions with the two boundaries, which are necessary to keep the particle inside the piston, namely $x_t \in [0,\lambda_t]$. Explicitly, $I_c$ suddenly changes $p$ into $2m\dot\lambda_t-p$ (or $-p$) once a collision at the right (or left) boundary occurs at time $t$. We emphasize that since the change of the momentum is essential in collision processes, our starting point is the underdamped \cite{Kurchan_1998,Cilliberto_2006,Kwon_2013} EOM (\ref{Langevin}) instead of the overdamped Langevin equation, which is simpler and is more frequently adapted in calculating work distributions in ST.%where the momentum degree of freedom is eliminated.

Provided that the collisions are elastic, the work functional in terms of a trajectory $\Gamma_t \equiv (x_t,p_t)$ in the phase space can be evaluated as \cite{Lua_2005}
\begin{equation}
W[\Gamma_t] = - \sum_{t \in C[x_t]} 2 \dot \lambda_t (p_{t^-} - m \dot \lambda_t),
\label{originalW}
\end{equation}
where $C[x_t] \equiv \{ t:x_t = \lambda_t, 0 \leq t \leq \tau \}$ is the set of collision time points for a trajectory $x_t$ in real space; $p_{t^-}$ is the momentum value at the time point immediately prior to $t$. One can see that the above work expression differs significantly from the usual one $W[x_t] = \int dt \partial_t V_t(x_t)$ \cite{Jarzynski_1997,Sekimoto_1998} in both the momentum dependence and the discrete summation rather than an integration.

To be specific, in the following we will focus on calculating the work distribution for the expansion process starting from a canonical ensemble, where the initial distributions of $x$ and $p$ are respectively $U(0,\lambda_0)$ (uniform distribution) and $N(0,m/\beta)$ (normal distribution). During the course the right boundary is driven according to an arbitrary protocol $\lambda_t$ and ends at $\lambda_\tau = 2\lambda_0$. %Now the model has been introduced, the problem is how to calculate the work distribution for an arbitrary given protocol $\lambda_t$.
Actually this is the model used in the famous BSE \cite{Hatano_1998,Broeck_2015,Szilard_1929,Leff_2003,Maruyama_2009,Dunkel_2014}.

\begin{figure}
\begin{center}
        \includegraphics[width=4.2cm, clip]{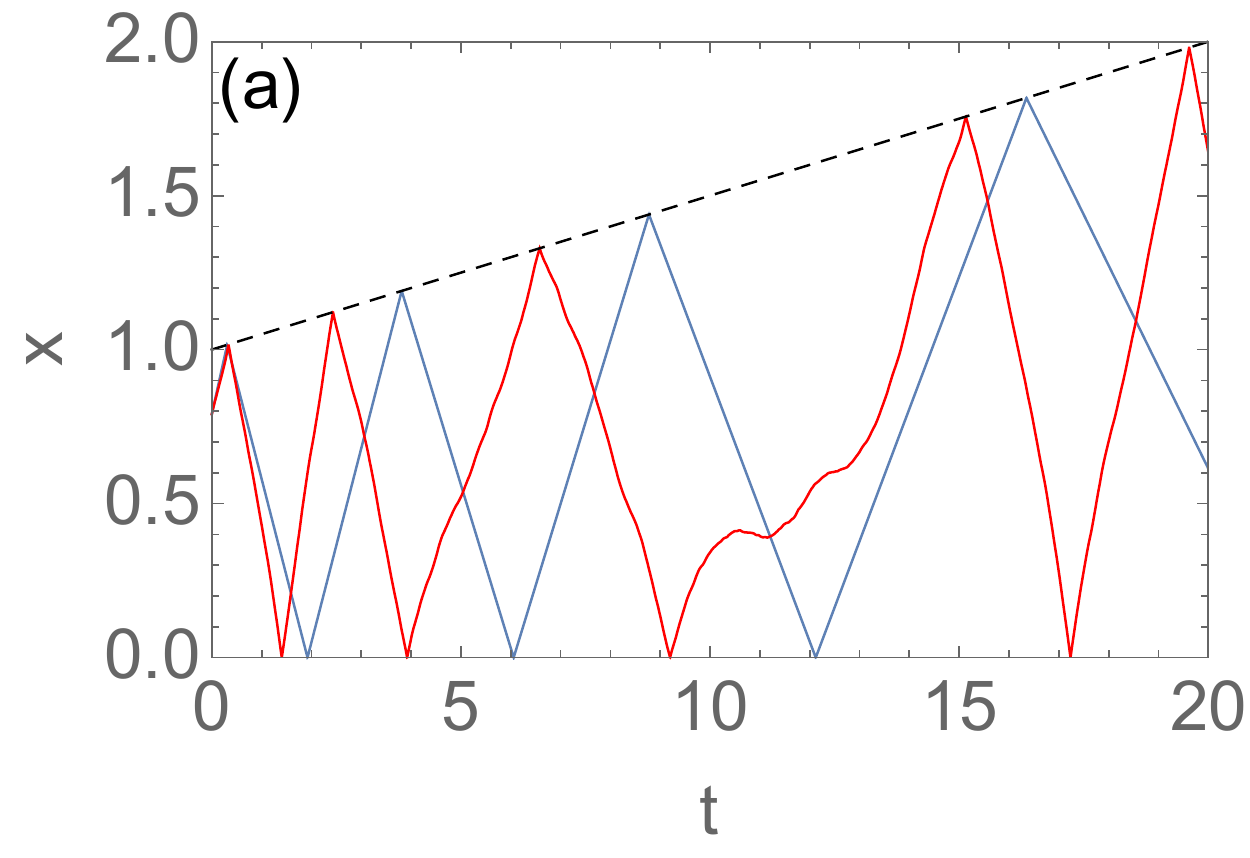}
        \includegraphics[width=4.2cm, clip]{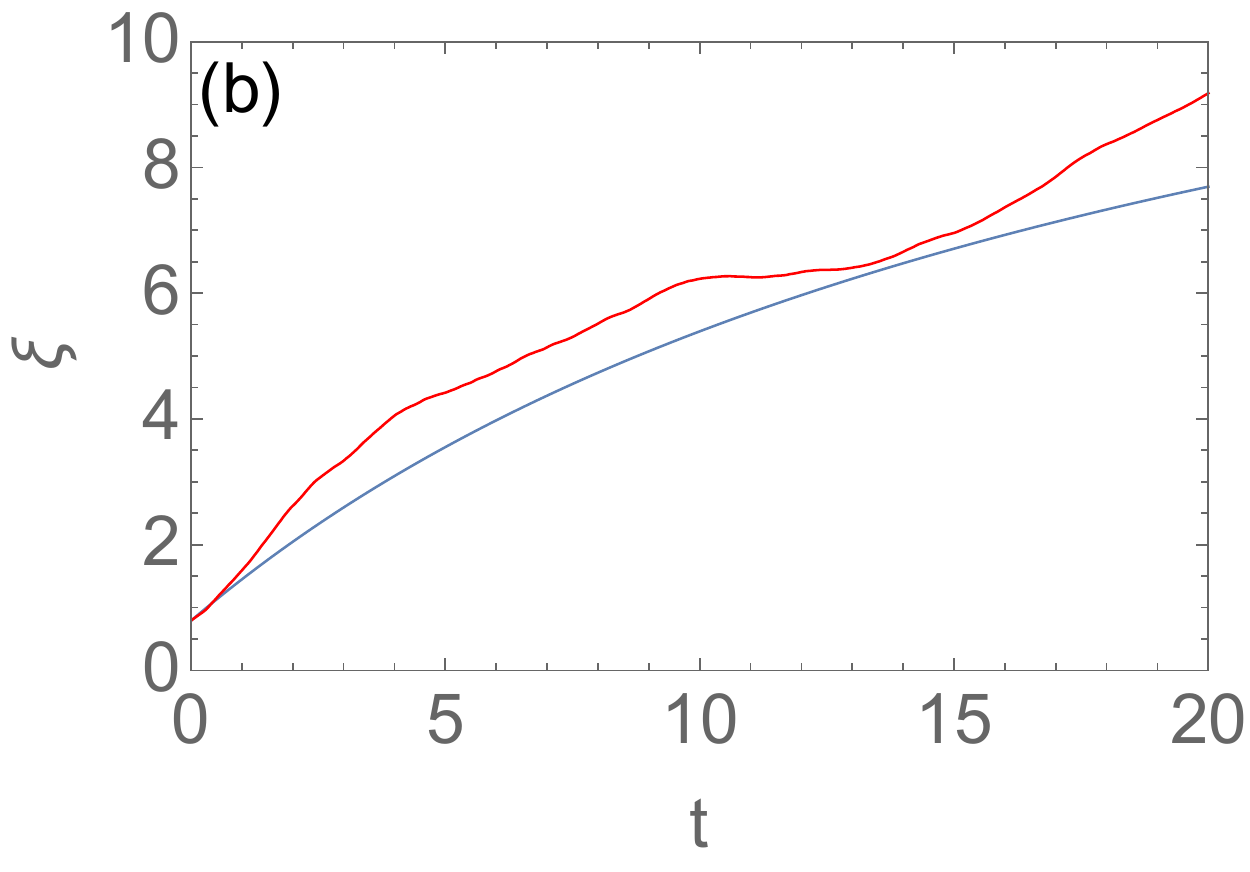} \ \
        \includegraphics[width=4.2cm, clip]{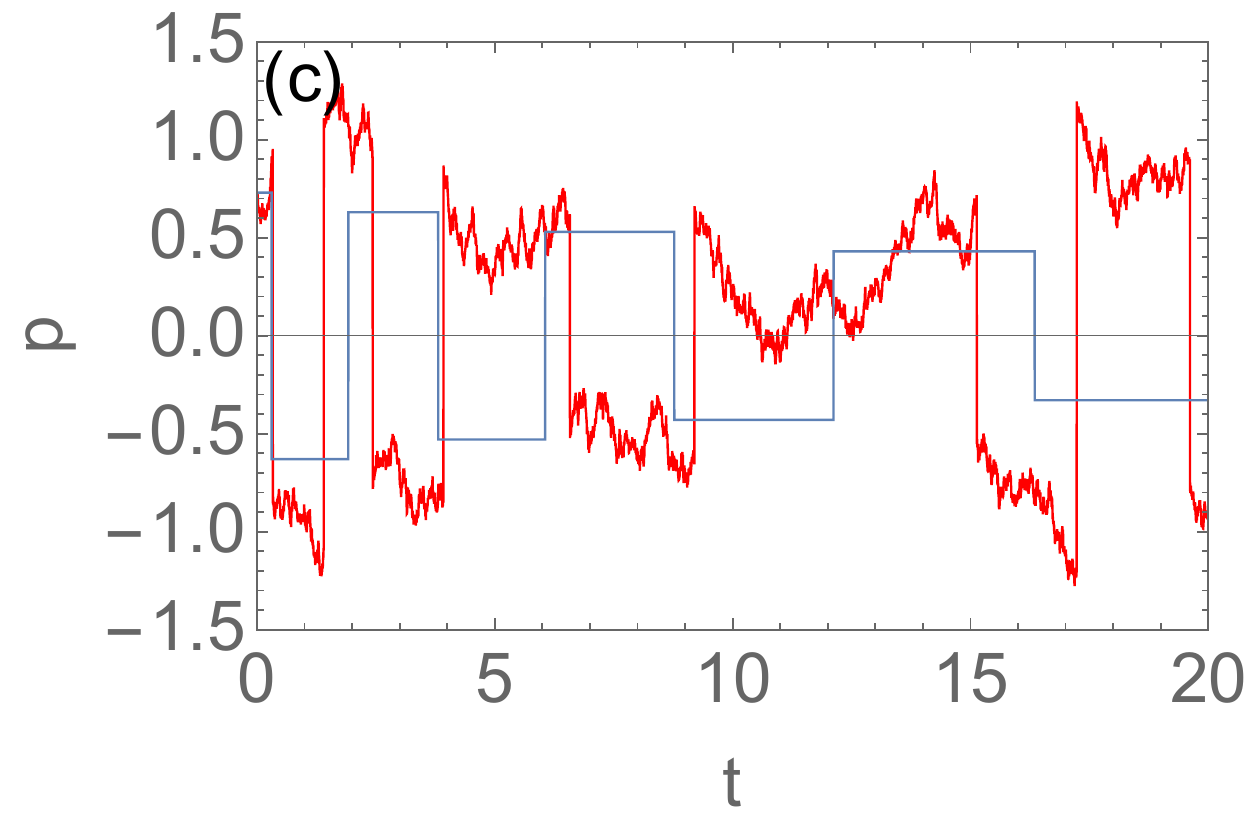}
        \includegraphics[width=4.2cm, clip]{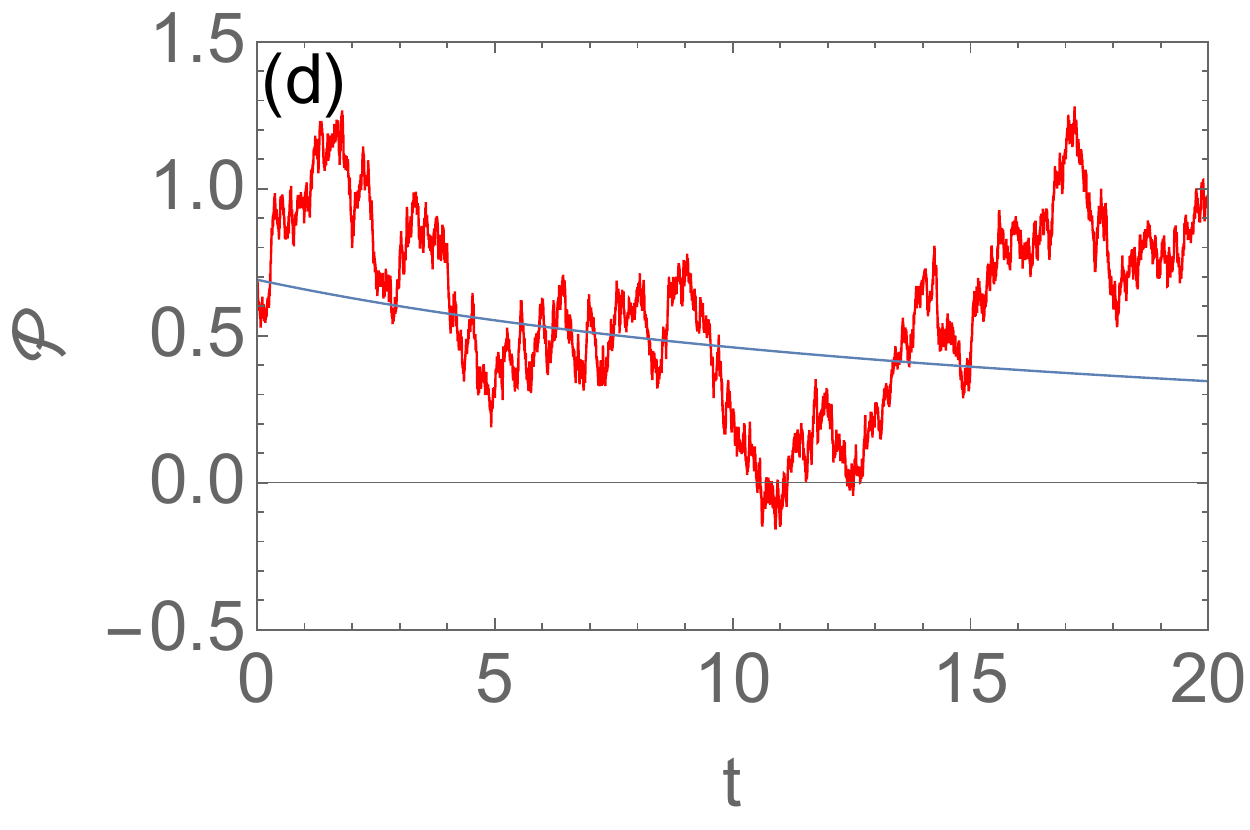}
      \end{center}
   \caption{Typical trajectories of  (a) position $x$ and (c) momentum $p$, as well as the new variables (b) $\xi$ and (d) $\mathcal{P}$ after transformation (\ref{vartrans}). The expansion protocol is the linear one, i.e., $\lambda_t = \lambda_0 (1 + t/\tau)$, where $\lambda_0=1$ and $\tau=20$, presented as the black dashed line in (a). All the blue curves correspond to the adiabatic process ($\gamma=0$), while the red ones correspond to the isothermal process with $\gamma=0.05$ and $\beta=1$.}\label{trajectory}
\end{figure}

\emph{Coordinate transformation and the Feynman-Kac equation.---} While the numerical simulation based on Eqs.~(\ref{Langevin}) and (\ref{originalW}) is straightforward, a direct analytical treatment seems to be hopeless, due to the difficulties caused by the time-dependent boundary condition and the collision term $I_c$. To eliminate these difficulties, we perform the following coordinate transformation \cite{SSM}
\begin{equation}
(-)^{\lfloor \xi \rfloor + 1} h(\xi) \equiv \frac{x}{\lambda_t} \;\; , \;\; \mathcal{P} \equiv (-)^{\lfloor \xi \rfloor} p + m \dot \lambda_t h(\xi),
\label{vartrans}
\end{equation}
where $h(\xi) \equiv 2 \lfloor (\xi + 1)/2 \rfloor - \xi$ with $\lfloor ... \rfloor$ being the Gauss floor function, the dimensionless quantity $\xi$ can be any value on the real axis. From Eq.~(\ref{vartrans}) it seems that the new coordinate $\xi$ can hardly be uniquely determined by $x$, but a one-to-one mapping between them can be indeed unambiguously established as long as we add the information of collision to $\xi$. We stipulate that $\xi$ crosses an integer every time a collision occurs. In particular, $\xi$ crosses an odd (even) integer once the particle collides with the right (left) boundary. Such correspondence relation (\ref{vartrans}) is illustrated schematically in Fig.~\hyperref[trajectory]{1}. It is found that both $\dot\xi$ and $\mathcal{P}$ are continuous functions of time, in the sense that they never jump. Thus we expect to construct a collision-free EOM with respect to the new variables $ \widetilde{\Gamma} \equiv (\xi,\mathcal{P})$, since they are continuous functions of time. After some calculations, we obtain the following new EOM in terms of $\xi$ and $\mathcal{P}$
\begin{equation}
\begin{split}
\dot \xi &= \frac{\mathcal{P}}{m \lambda_t},\\
\mathcal{\dot P} &= \left(\gamma \dot \lambda_t + m \ddot \lambda_t\right) h(\xi) - \left(\frac{\gamma}{m} + \frac{\dot \lambda_t}{\lambda_t}\right) \mathcal{P} + \sqrt{\frac{2 \gamma}{\beta}} \eta_t.
\end{split}
\label{EOM}
\end{equation}
Correspondingly, the work functional in terms of the new variables reads
\begin{equation}
W[\widetilde{\Gamma}_t] = - \int^\tau_0 dt \frac{\mathcal{P}^2_t \dot \lambda_t}{m \lambda_t} \left[h'(\xi_t) + 1 \right],
\label{workfunctional}
\end{equation}
where $h'(\xi) + 1$ is a compact form of $2 \sum_{k \in \mathbb{Z}} \delta(\xi - 2k - 1)$, and obviously this work functional (\ref{workfunctional}) cannot be directly obtained from the usual microscopic definition of work $W[x_t] = \int dt \partial_t V_t(x_t)$ \cite{Jarzynski_1997,Sekimoto_1998}.

\begin{figure}
\begin{center}
        \includegraphics[width=8cm, clip]{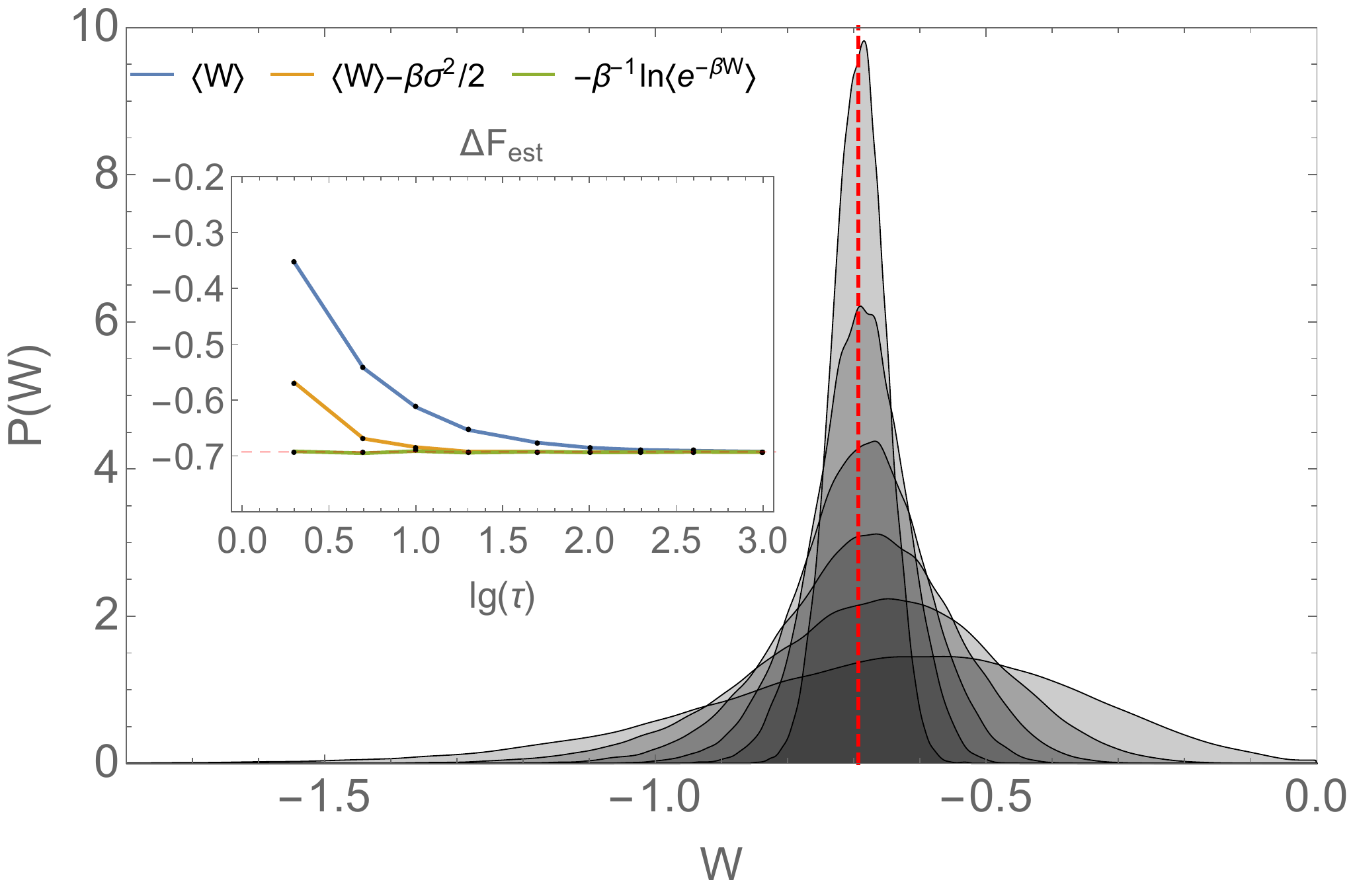}
      \end{center}
   \caption{Work distribution functions for the uniform expansion protocol with $\tau = 20,50,100,200,400,1000$ obtained from stochastic simulations, where the $P(W)$ curve with sharper peak corresponds to larger $\tau$ (similar results were obtained for a breathing HO in \cite{Jarzynski_1997b}). The vertical red dashed line marks the position of $\Delta F=\beta^{-1}\ln2$. Inserted figure shows the numerical  estimation of the free energy difference $\Delta F_{\mathrm{est}}$ based respectively on the mean work $\langle W\rangle$ (blue line), the linear response correction $\langle W\rangle-\beta\sigma^2/2$ (orange line) and the JE $-\beta^{-1}\ln\langle e^{-\beta W}\rangle$ \cite{Hendrix_2001} for nine different uniform expansion processes, with $\tau = 2,5,10,20,50,100,200,400,1000$. The horizontal red dashed line is the theoretical free energy difference while the dots are the simulation results. Here $\lambda_0=1$, $\beta=1$ and $\gamma=1$ are all fixed.}\label{simulation}
\end{figure}

To check the correctness of such coordinate transformation, we carry out numerical simulations based on the new EOM (\ref{EOM}) and the work functional (\ref{workfunctional}). The results are presented in Fig.~\hyperref[simulation]{2}, which strongly suggest the validity of the Jarzynski equality (JE) and the asymptotic Gaussian type of work distribution for large $\tau$, which has been analytically demonstrated for generic overdamped Langevinian systems with smooth potentials \cite{Speck_2004}. Further examinations confirm the validity of the coordinate transformation \cite{SSM}.

With these relations, we can write down the Feynmann-Kac equation (FKE) \cite{Hummer_2001,Speck_2011,Kwon_2013}, which determines the time evolution of the phase point distribution weighted by a parametric exponential work factor. The FKE is obtained as \cite{SSM}
\begin{equation}
\partial_t \rho_s = \mathcal{L}[\lambda_t] \rho_s + s \frac{\mathcal{P}^2 \dot \lambda_t}{m \lambda_t} [h'(\xi) + 1] \rho_s,
\label{FKE}
\end{equation}
where $\rho_s = \rho_s (\xi,\mathcal{P},t)$ is related to the joint distribution function $\rho (\xi,\mathcal{P},W,t)$ by a Laplace transform $\rho_s \equiv \int^{+ \infty}_{-\infty} dW \rho (\xi,\mathcal{P},W,t) e^{-sW}$, and the linear operator $\mathcal{L}[\lambda_t]$ is defined as
\begin{equation}
\begin{split}
\mathcal{L}[\lambda_t] \equiv &-\frac{\mathcal{P}}{m \lambda_t} \partial_\xi + \left[\left(\gamma \dot \lambda_t + m \ddot \lambda_t\right) h(\xi) \partial_{\mathcal{P}}\right.  \\
&+ \left.\partial_{\mathcal{P}} \left(\frac{\gamma}{m} + \frac{\dot \lambda_t}{\lambda_t}\right)\mathcal{P}\right] + \frac{\gamma}{\beta} \partial^2_{\mathcal{P}}.
\end{split}
\end{equation}
Once we solve Eq.~(\ref{FKE}), we can immediately obtain the generating function $\psi_s (t)$ of the work distribution by integrating out $\xi$ and $\mathcal{P}$, namely $\psi_s (t) \equiv \langle e^{-sW} \rangle = \int d\xi d\mathcal{P} \rho_s (\xi,\mathcal{P},t)$. The generating function $\psi_s (t)$ provides an alternative way to get access to the properties of the work distribution function \cite{Speck_2011}, so the central problem is to solve the FKE (\ref{FKE}).

\emph{Frequent collision approximation and the reduced Feynman-Kac equation---}Unfortunately, a general exact solution of the FKE (\ref{FKE}) is difficult to obtain, due to the complexities arising from both the number of variables and the non-analycity of the expressions ($h(\xi)$). In fact, besides the driven overdamped Brownian HO \cite{Jarzynski_1999,Kim_2014}, the V-potential \cite{Rybov_2015}, and the logarithmic-harmonic potential \cite{Rybov_2013,Rybov_2015}, the only analytically solvable model in ST so far seems to be the breathing overdamped Brownian HO \cite{Speck_2011,Kwon_2013}. Even for such a model, an exact solution is usually unavailable unless the initial distribution is Gaussian. %The difficulty of obtaining a general exact solution of Eq.~(\ref{FKE}) can be further comprehended from the result of another former work that calculated the exact work distribution function for the very special case $\gamma = 0$ (adiabatic) and $\ddot \lambda_t = 0$ (uniform speed of the piston) \cite{Lua_2005}. Owing to the unique work production mechanism (collision) in piston systems, even the work distribution function of such simplest case ($\gamma = 0$, $\ddot \lambda_t = 0$) turned out to be very complicated (see Eqs.~(13-16) in Ref.~\cite{Lua_2005}). Hence, one can hardly anticipate an exact generalization to nonzero $\gamma$ and arbitrary $\lambda_t$.

Accordingly, we need to make further approximations to obtain analytic results in certain interesting regimes. Remember that one of the difficulties comes from the discreteness of collisions, and the work accumulates more and more continuously as the collision frequency increases. This is the case in the high temperature limit for a given protocol, or equivalently, in the slow limit of the protocol at any finite temperature. A paradigmatic example to illustrate this subtlety is the work distribution for the quasistatic adiabatic expansion processes of an ideal gas \cite{Crooks_2007}, which can be exactly reproduced by the universal work distribution function in Ref.~\cite{Lua_2005} via smoothing out the local oscillations caused by the discreteness of collisions. Inspired by this, we can similarly try to flatten the rapidly oscillating parts $h(\xi)$ in the FKE (\ref{FKE}). In fact, it is feasible to construct a reduced partial differential equation only in terms of $\mathcal{P}$ via integrating out the position-like variable $\xi$ under this approximation, which is completely in contrast to the conventional overdamp Langevin dynamics where the position instead of the momentum is kept. The reduced Feynman-Kac equation (RFKE) in this case is
\begin{equation}
\partial_t \varrho_s = \partial_{\mathcal{P}} \left[\left(\frac{\gamma}{m} + \frac{\dot \lambda_t}{\lambda_t}\right)\mathcal{P} + \frac{\gamma}{\beta} \partial_{\mathcal{P}}\right] \varrho_s + s \frac{\mathcal{P}^2 \dot \lambda_t}{m \lambda_t} \varrho_s,
\label{RFKE}
\end{equation}
where $\varrho_s = \varrho_s (\mathcal{P},t) \equiv \int d\xi \rho_s (\xi,\mathcal{P},t)$ is the $\mathcal{P}$ marginal distribution function weighted by a parametric exponential work factor. The validity of the RFKE can be checked self-consistently \cite{SSM}. We emphasize that the RFKE (\ref{RFKE}) is merely an approximated equation valid for sufficiently slow expanding. %Mathematically speaking, we actually neglect the two integrals $\int d\xi h(\xi) \partial_{\textcolor{red}{\mathcal{P}}} \rho_s$ and $\int d\xi h(\xi) \partial_\xi \rho_s$ in the derivation of the RFKE, which is reasonable when the width of $\rho_s$ is much larger than the period of function $h(\xi)$. The more frequently the collisions, the more diffusive in $\xi$ dimension is $\rho_s$, hence leading to the higher precision of the RFKE.

\emph{Asymptotic behavior and protocol optimization in the linear response regime.---}Thanks to the similarity between the RFKE (\ref{RFKE}) (as well as its associated work functional) and the overdamped FKE for a breathing HO \cite{Speck_2011,Kwon_2013}, we can further simplify Eq.~(\ref{RFKE}) into a set of ordinary differential equations (ODEs) by utilizing the technique developed in dealing with the breathing HO model \cite{Speck_2011,Kwon_2013}. The key point of the technique is the Gaussian ansatz that the solution takes the form $\varrho_s (\mathcal{P},t) = \sqrt{\frac{[\psi_s (t)]^3}{2 \pi \phi_s (t)}} e^{-\frac{\mathcal{P}^2 \psi_s (t)}{2 \phi_s (t)}}$ \cite{Footnote}. In this manner, the RFKE (\ref{RFKE}) is equivalent to
\begin{equation}
\dot \psi_s = \frac{s \dot \lambda_t}{m \lambda_t} \phi_s ,
\dot \phi_s = -2\left(\frac{\gamma}{m} + \frac{\dot \lambda_t}{\lambda_t}\right)\phi_s + \frac{2 \gamma}{\beta} \psi_s + \frac{3 s \dot \lambda_t}{m \lambda_t} \frac{\phi^2_s}{\psi_s},
\label{ODEs}
\end{equation}
with the initial conditions $\psi_s (0) = 1$ and $\phi_s (0) = m/\beta$.
%To overcome such difficulty, we focus on the linear response regime, so that a Gaussian initial condition is a good approximation.

\begin{figure}
\begin{center}
        \includegraphics[width=7cm, clip]{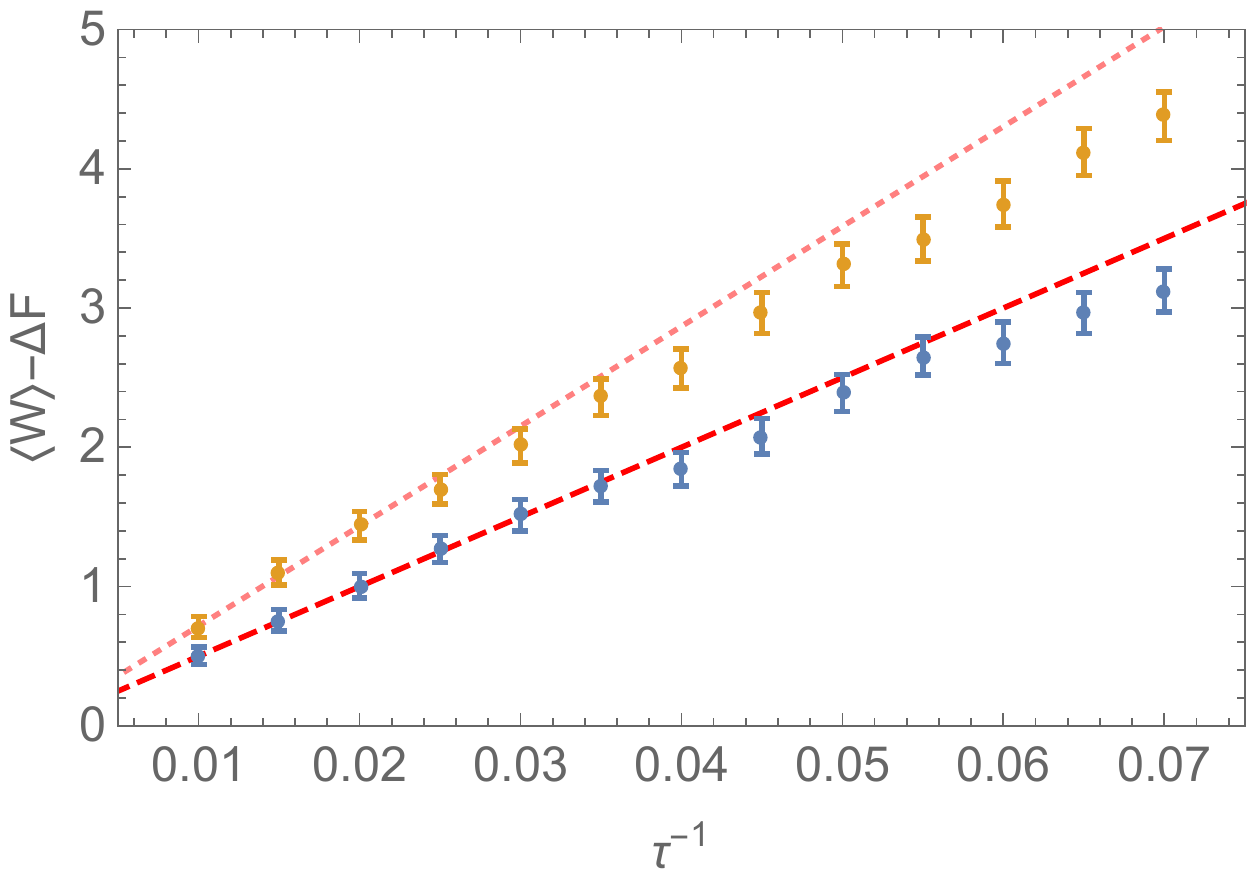}
     \end{center}
   \caption{$\langle W\rangle - \Delta F$ versus $\tau^{-1}$ for the uniform and the sine expansion protocols in the linear response regime, obtained by numerical stochastic simulations based on the original EOM (\ref{Langevin}) as well as the original work expression (\ref{originalW}) (blue and yellow dots), and the theoretical prediction (\ref{FDR2}) (red dashed and dotted lines). The parameters are $\gamma = 1$ and $\beta^{-1} = 100$, and the error bar denotes twice the standard deviation of the mean. One can see good agreement for sufficiently small $\tau^{-1}$. %when $\tau^{-1}<0.035$.
   } \label{fig3}
\end{figure}

To proceed analytically, we further confine ourselves in the linear response regime, where $\alpha \equiv  m \dot \lambda_t/\gamma \lambda_t\ll1$ and Eq.~(\ref{ODEs}) can be solved perturbatively. To perform perturbative analysis, we introduce another function $g_s (t) \equiv \frac{\beta}{s} \frac{d \ln \psi_s}{d \ln \lambda_t}$, thus $\psi_s$ can be evaluated in terms of $g_s$ through $\psi_s (\tau) = \exp [\frac{s}{\beta} \int^\tau_0 dt \frac{\dot \lambda_t}{\lambda_t} g_s (t)]$. Now the problem is to solve for $g_s (t)$. The nonlinear ODE that governs the time evolution of $g_s (t)$ is found to be a Riccati equation
\begin{equation}
\dot g_s = \frac{2 \dot \lambda_t}{\lambda_t } g_s \left(\frac{s}{\beta} g_s - 1\right) - \frac{2 \gamma}{m}\left(g_s - 1\right),
\;\;\;\; g_s(0) = 1.
\end{equation}
In the sense of perturbation, $g_s$ should be expanded as $1 + g^{(1)}_s + g^{(2)}_s + ...$, where the magnitude of $g^{(k)}_s$ is $O(\alpha^k)$. %It is notable that $g_s (t) = 1$ is an exact solution when $s = \beta$, implying $\psi_\beta (\tau) = \lambda_\tau / \lambda_0$ and finally leading to the JE again.
In the linear response regime, we have $g_s \approx 1+g^{(1)}_s$, according to which we expect the system to obey the FDR. In fact, we obtain $g^{(1)}_s = \frac{m \dot \lambda_t}{\gamma \lambda_t} (\frac{s}{\beta}-1)$, thus the generating function should be
\begin{equation}
\psi_s (\tau) = \exp\left[\frac{s}{\beta} \ln \frac{\lambda_\tau}{\lambda_0} + \frac{s}{\beta}\left(\frac{s}{\beta} - 1\right)\frac{\gamma}{m} \int^\tau_0 dt \alpha^2 + O(\alpha^2)\right].
\end{equation}
This expression indicates that the corresponding work distribution is Gaussian with the mean $\langle W \rangle = -\beta^{-1} \ln (\lambda_\tau / \lambda_0) + \frac{\gamma}{\beta m} \int^\tau_0 dt \alpha^2$ and the variance $\sigma^2_W = \frac{2 \gamma}{\beta^2 m} \int^\tau_0 dt \alpha^2$. Since the free-energy difference is $\Delta F = -\beta^{-1} \ln (\lambda_\tau / \lambda_0)$, we verify the first prediction of the FDR \cite{Hendrix_2001}: $\langle W \rangle - \Delta F = \frac{1}{2} \beta \sigma^2_W $.
If we define the protocols $\lambda_t = \Lambda(t/\tau)$ with different $\tau$ as one class, then for a given class $\Lambda(u)$ $(0 \leq u \leq 1)$, the deviation of the mean work from the free-energy difference will be inversely proportional to $\tau$ \cite{Bechhoefer_2014}
\begin{equation}
\langle W \rangle - \Delta F = K \tau^{-1} \; ,
\label{FDR2}
\end{equation}
where the coefficient $K = \frac{m}{\beta \gamma} \int^1_0 du [\chi' (u)]^2 $, $\chi(u) \equiv \ln \frac{\Lambda(u)}{\Lambda(0)}$. For a linear (sine) protocol $\chi(u)=\ln(1+u)$ ($\chi(u)=\ln\left[2\sin\frac{\pi}{3}\left(u+\frac{1}{2}\right)\right]$), we have $K = \frac{m}{2\beta \gamma}$ ($K=\frac{(3\sqrt{3}-\pi)\pi m}{9\beta\gamma}$). This is another prediction of FDR in the linear response regime, and is numerically verified (see Fig.~\ref{fig3}). %for the uniform speed expansion protocol.
So far, we have analytically demonstrated that all these asymptotic behaviors of the work distribution of the expanding isothermal piston system share the same features with those of conventional overdamped Langevin systems \cite{Jarzynski_1997b,Speck_2004,Hendrix_2001}, and obey FDRs. However, we again emphasize that the approaches used to deal with the systems with smooth potentials are essentially inapplicable to the piston system. %Also, in the ``thermal wall model" used in Refs. \cite{Hatano_1998,Baule_2006,Engel_2013,Broeck_2015}, the FDRs cannot be verified analytically in the linear response regime.
So a distinct method for the piston system is developed here.

Since we have obtained the mean work expression (\ref{FDR2}) analytically, we can also investigate the optimization problem in the linear response regime. Particularly, we are interested in the maximum mean work extraction from the heat bath for a given time interval $[0,\tau]$ \cite{Seifert2007,Engel_2008,Marcus_2014}, because the optimal work protocol of the BSE is a very important but unsolved problem. For the expansion process in a BSE cycle, the boundary condition can be rewritten as $\chi(0) = 0$ and $\chi(1) = \ln 2$. To maximize the mean work extraction, we only have to minimize the coefficient $K$ as a functional of $\chi(u)$. The variation of $K$ in terms of $\chi(u)$ gives a simple equation $\chi'' (u) = 0$, implying that $\chi(u) = u \ln 2$ or $\lambda_t = \lambda_0 2^{t/\tau}$ is the optimal protocol that makes $K$ reach its minimum $\frac{m}{\beta \gamma} \ln^2 2$. Starting from Eq. (\ref{RFKE}), the same result can be obtained from the thermodynamic length $\mathcal{L}$ via $K=\mathcal{L}^2$ \cite{Crooks_2012}, where $\mathcal{L}=\int^{\lambda_\tau}_{\lambda_0}d\lambda\sqrt{\zeta}$ with $\zeta=\frac{m}{\beta\gamma\lambda^2}$ being the thermodynamic metric for the piston system.

It is worth mentioning that we may also analyze the optimization problem based on Eq.~(\ref{RFKE}) without doing perturbative expansions. The optimal protocol %by following the procedure of Ref. \cite{Seifert2007}
turns out to be similar to that of the breathing HO \cite{Seifert2007}, %which contains sudden jump at both the initial and the final point. However, the analytic expressions are completely different (see supplemental material). Unfortunately, the sudden jump feature contradicts with the assumption leading to Eq.~(\ref{RFKE}), except
and in the linear response regime, %where the jump becomes vanishingly small and
the exponential optimal protocol $\lambda_t=\lambda_02^{t/\tau}$ can be reproduced. %\cite{General}.
%This result might be a little surprising, because it is usually anticipated that the uniform speed expansion should be optimal. In fact, the coefficient $K$ corresponding to the linear protocol $\frac{m}{2 \beta \gamma}$ is indeed larger than that of the optimal protocol ($\ln^2 2 \approx 0.480 <0.5$).

\emph{Conclusion---}Previously nonequilibrium thermodynamics in the isothermal piston system can only be studied numerically and few insights can be gained from the numerical results \cite{Talkner}. In this Letter, by performing a coordinate transformation, we find that the EOM in the new coordinate corresponds to a collision-free stochastic diffusive system %with a time-dependent periodic potential
in the full space. We have derived the exact %universal
FKE and simplified it into a single-variable RFKE under the frequent collision approximation. %The simplified equation looks similar to the overdamped FKE for the well-investigated breathing HO. By making use of the exact solution of breathing HO \cite{Speck_2011,Kwon_2013}, we solved the RFKE in a perturbative manner in the linear response regime, and obtain the work distribution analytically for an arbitrary protocol $\lambda_t$. With the new method
By solving the RFKE perturbatively, we not only demonstrate the Gaussian asymptotic behavior of the work distribution and the validity of the FDRs in the piston system, but also obtain the optimal work extraction protocol $\lambda_t = \lambda_0 2^{t/\tau}$ of the BSE in the linear response regime. %Some of our analytical results are examined by numerical stochastic simulations, which show good agreements.
%We suggest the methodology developed in this Letter can be generalized to deal with other rigid wall pontential stochastic thermodynamic systems, which are nearly unexplored so far. The optimization of the BSE beyond the linear response regime may also be a challenging subject to investigate.
Our study is complementary to previous studies of ST in systems with smooth potentials. By extending the studies of ST to the conceptually simplest and paradigmatic model in traditional thermodynamics ---the isothermal piston system, we bridge the long-standing gap in the development of ST. %The new method could be used in many other interesting studies, e.g., the Carnot efficiency at the maximum power based on a piston system, and the optimization of the BSE beyond the linear response.

\textbf{Acknowledgments} H. T. Q. gratefully acknowledges support from the National Science Foundation of China under grants 11375012, 11534002, and The Recruitment Program of Global Youth Experts of China. Y. H. L. is supported by National Natural Science Foundation of China (Grant No. 11375093), MOST 2013CB922000 of the National Key Basic Research
Program of China. Z. G. is supported by MEXT scholarship.

\bibliography{GZP_references}

\begin{thebibliography}{67}
\bibitem[{Foo()}]{Footnote}
\bibinfo{note}{However, according to the definition of $\mathcal{P}$
  (\ref{vartrans}), since the real momentum $p \thicksim N(0,m/\beta)$
  initially, the initial distribution of $\mathcal{P}$ is not rigorously
  Gaussian, unless $\dot \lambda_0$ is vanishingly small. But since we work in
  the frequent-collision regime with $\dot\lambda_t\ll(\beta m)^{-1/2}$, a
  Gaussian initial condition is a good approximation.}
\end{thebibliography}

\clearpage
\begin{center}
\textbf{\large Supplemental Materials}
\end{center}
\setcounter{equation}{0}
\setcounter{figure}{0}
\setcounter{table}{0}
\makeatletter
\renewcommand{\theequation}{S\arabic{equation}}
\renewcommand{\thefigure}{S\arabic{figure}}
\renewcommand{\bibnumfmt}[1]{[S#1]}
\begin{center}
Here we provide the detailed derivations of Eqs.~(3), (6) and (8) in the main text, and other useful information.
\end{center}

\appendix

\section{Details of the coordinate transformation Eq.~(3)}
Since the main difficulty of the problem comes from the time-dependent boundary condition, we first use the Lagrange picture to change the original position coordinate $x_t$ into $\zeta_t\equiv x_t/\lambda_t$, so that the range of $\zeta_t$ ($[0,1]$) is time-independent. This technique has already been used by Nakamura et al. \cite{Nakamura2011,Nakamura2012}, but for dealing with the quantum piston system. Another advantage of using $\zeta_t$ is that $\dot \zeta_t$ either changes smoothly or have a sudden change via sign inversion. This property results from the simple fact that in the inertial frame where a rigid wall is static, a small ball always inverts its velocity after elastically colliding with the wall.

\begin{figure}
\begin{center}
        \includegraphics[width=7cm,clip]{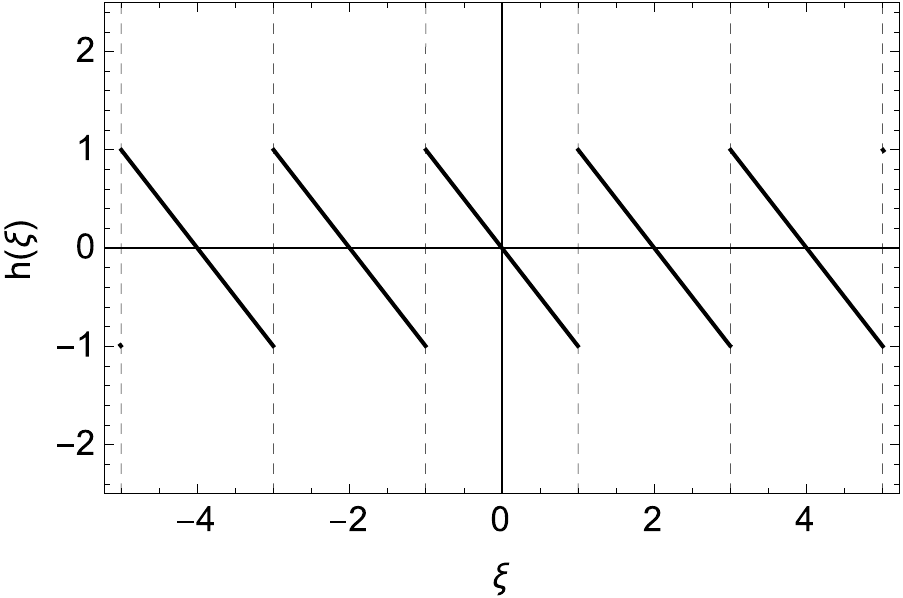}
   \end{center}
   \caption{Periodic function $h(\xi)\equiv 2\lfloor (\xi+1)/2\rfloor-\xi$.}
\label{fig1}
\end{figure}

So far the problem is still a bounded one and the singular term in the equation of motion of $\zeta$ is not eliminated yet. To find out a position-like variable $\xi$ which evolves continuously, we relate $\dot\xi$ to $\dot\zeta$ by multiplying $-1$ right after each collision. Notice that a collision occurs when and only when $\zeta=0$ or $1$, so the following transformation should fit our requirement
\begin{equation}
\xi_t=\zeta_0+\int^t_0 dt' (-)^{N_c[\zeta_{t'};t]}\dot \zeta_{t'},
\label{xivszeta}
\end{equation}
where $N_c[\zeta_{t'};t]\in\mathbb{N}$ counts the number of collisions during $[0\,,t]$ along the trajectory $\zeta_{t'}$. What's more, after the transformation $\xi_t$ ranges from $-\infty$ to $+\infty$, so the problem becomes boundless. To write down Eq.~(\ref{xivszeta}) more elegantly, we use the inverted map
\begin{equation}
\zeta_t=(-)^{\lfloor\xi_t\rfloor+1}h(\xi_t),
\label{zetavsxi}
\end{equation}
where $\lfloor\cdot\rfloor$ is the Gauss floor function and $h(\xi)\equiv2\lfloor(\xi+1)/2\rfloor-\xi$ is a periodic function with period 2 (see Fig.~\ref{fig1}). Though the inverse map is not monotonic, $\xi_t$ can be indeed uniquely determined by $\zeta_t$ if we further provide the discrete information that $\xi_t$ crosses an integer once a collision occurs, as has been mentioned in the main text.

Now we have obtained a position-like variable $\xi_t$ with continuous derivative $\dot\xi_t$, so that $\mathcal{P}_t\equiv m\lambda_t\dot\xi_t$ must also be a continuously evolving variable related to the momentum.  By making use of the relation $\dot\xi_t=(-)^{\lfloor\xi_t\rfloor}\dot\zeta_t$ which follows Eqs.~(\ref{xivszeta}) and (\ref{zetavsxi}), we can connect $\mathcal{P}_t$ to the original momentum $p_t=m\dot x_t$ as follows
\begin{equation}
\begin{split}
\mathcal{P}_t&=m\lambda_t\dot\xi_t=(-)^{\lfloor\xi_t\rfloor}m\lambda_t\dot\zeta_t=(-)^{\lfloor\xi_t\rfloor}m \left(\dot x_t-\dot\lambda_t\zeta_t\right)\\
&=(-)^{\lfloor\xi_t\rfloor}m \left[\dot x_t-\dot\lambda_t (-)^{\lfloor\xi_t\rfloor+1}h(\xi_t)\right] \\
&=(-)^{\lfloor\xi_t\rfloor}p_t+m\dot\lambda_t h(\xi_t).
\end{split}
\label{pP}
\end{equation}
Finally we complete the coordinate transformation from $(x_t,p_t)$ to $(\xi_t,\mathcal{P}_t)$
\begin{equation}
(-)^{\lfloor \xi_t \rfloor + 1} h(\xi_t) \equiv \frac{x_t}{\lambda_t},  \;\;\;\; \mathcal{P}_t \equiv (-)^{\lfloor \xi_t \rfloor} p_t + m \dot \lambda_t h(\xi_t).
\end{equation}
In terms of the new variables $\xi_t$ and $\mathcal{P}_t$, the equation of motion must be nonsingular, namely there should be no longer collision terms manifesting as delta functions.

\section{Details of the derivation of the FKE Eq.~(6)}
Let's first figure out the equation of motion in terms of the new coordinates. Notice that when there is no collision, the particle simply undergoes free Brownian motion, of which the dynamics is described by the Langevin equation
\begin{equation}
\dot x_t=\frac{p_t}{m},\;\;\;\;\; \dot p_t=-\frac{\gamma}{m}p_t+\sqrt{\frac{2\gamma}{m}}\eta_t,
\end{equation}
where $\eta_t$ is the standard Wiener process. By substituting $p_t$'s expression in terms of $\mathcal{P}_t$ and $\xi_t$ into the second equation above, we obtain
\begin{equation}
\begin{split}
&(-)^{\lfloor\xi_t\rfloor}\left[\mathcal{\dot P}_t-m\ddot\lambda_t h(\xi_t)+m\dot\lambda_t\dot\xi_t \right] \\
=&-\frac{\gamma}{m}(-)^{\lfloor\xi_t\rfloor}\left[\mathcal{P}_t-m\dot\lambda_th(\xi_t) \right]+\sqrt{\frac{2\gamma}{\beta}}\eta_t.
\end{split}
\label{eqP}
\end{equation}
Here we use the fact that $h'(\xi_t)=-1$ and $\lfloor\xi_t\rfloor$ stays unchanged when no collision occurs ($\xi_t$ varies between two adjacent integers). Notice that the white noise $\eta_t$ is symmetric and with zero mean, we have $(-)^{\lfloor\xi_t\rfloor}\eta_t=\eta_t$. By replacing $\dot\xi_t$ with $\mathcal{P}_t/m\lambda_t$ in Eq.~(\ref{eqP}), we obtain
\begin{equation}
\mathcal{\dot P}_t=\left(\gamma\dot\lambda_t+m\ddot\lambda_t\right)h(\xi_t)-\left(\frac{\gamma}{m}+\frac{\dot\lambda_t}{\lambda_t}\right)\mathcal{P}_t+\sqrt{\frac{2\gamma}{\beta}}\eta_t.
\label{EOM}
\end{equation}
Combining with $\dot\xi_t=\mathcal{P}/m\lambda_t$, we finally get the equation of motion in the main text (Eq.~(4)).

Next we derive the work functional. The work functional in terms of $(x_t,p_t)$ has already been given by Eq.~(2) in the main text. To rewrite it in terms of $(\xi_t,\mathcal{P}_t)$, we first mention that, immediately before a collision $t=t_c-\epsilon$, ($t_c\in C[x_t]$ so that $\xi_t$ is extremely close to an odd integer), according to Eq.~(\ref{pP}), we have
\begin{equation}
(-)^{\lfloor\xi_t\rfloor}\mathcal{P}_t=p_t-m\dot\lambda_t.
\end{equation}
Here we have used the continuity property of $\mathcal{P}_t$. While we cannot decide the sign of $(-)^{\lfloor\xi_t\rfloor}$, we definitely know that $p_{t_c-\epsilon}-m\dot\lambda_{t_c}>0$ (otherwise no collision occurs), so that $|\mathcal{P}_{t_c}|=p_{t_c-\epsilon}-m\dot\lambda_{t_c}$. Also, we have $C[x_t]=C[\xi_t]=\{t:\xi_t\in2\mathbb{Z}+1,0\le t\le\tau\}$, thus
\begin{equation}
\begin{split}
W[\widetilde{\Gamma_t}]&=-\sum_{t\in C[\xi_t]}2\dot\lambda_t|\mathcal{P}_t| \\
&=\int^\tau_0 dt 2\dot\lambda_t|\mathcal{P}_t| \sum_{t_c\in C[\xi_t]} \delta(t-t_c).
\end{split}
\end{equation}
Using the property of delta function $|f'(x)|\delta(f(x))=\sum_{x_0\in Z[f(x)]}\delta(x-x_0)$, where $Z[f(x)]\equiv\{x:f(x)=0,x\in\mathbb{R}\}$ is the set of zero points, we can rewrite the above expression as follows
\begin{equation}
\begin{split}
W[\widetilde{\Gamma_t}]&=\int^\tau_0 dt 2\dot\lambda_t|\mathcal{P}_t| \sum_{\xi_c\in 2\mathbb{Z}+1} |\dot\xi_t|\delta(\xi_t-\xi_c) \\
&=\int^\tau_0 dt\frac{2\mathcal{P}^2_t\dot\lambda_t}{m\lambda_t} \sum_{\xi_c\in 2\mathbb{Z}+1}\delta(\xi_t-\xi_c).
\end{split}
\label{functional}
\end{equation}
After replacing $2\sum_{\xi_c\in2\mathbb{Z}+1}\delta(\xi_t-\xi_c)$ with its compact form $h'(\xi)+1$, we finally get Eq.~(5) in the main text.

With the Langevin equation and the work functional in hand, we can thus construct the FKE. First, we apply the It\^o's lemma to $\delta(\xi_t-\xi)\delta(\mathcal{P}_t-\mathcal{P})$ \cite{Sekimoto2010}. According to the equation of motion for $(\xi_t,\mathcal{P}_t)$, we obtain the Fokker-Planck equation $\partial_t\rho(\widetilde\Gamma,t) = \mathcal{L}[\lambda_t]\rho(\widetilde\Gamma,t)$, where the expression of the generator $\mathcal{L}[\lambda_t]$ has been given by Eq.~(7) in the main text. Second, we modify the generator by adding $-s w(\widetilde\Gamma,t)$, where $W[\widetilde{\Gamma}_t]=\int^\tau_0 dt w(\widetilde\Gamma_t,t)$ \cite{Sekimoto2010,Speck2011}, to finally obtain the Feynman-Kac equation (Eq.~(6) in the main text).

\section{Independent test of the coordinate transformation}

\begin{figure}
\begin{center}
        \includegraphics[width=7cm,clip]{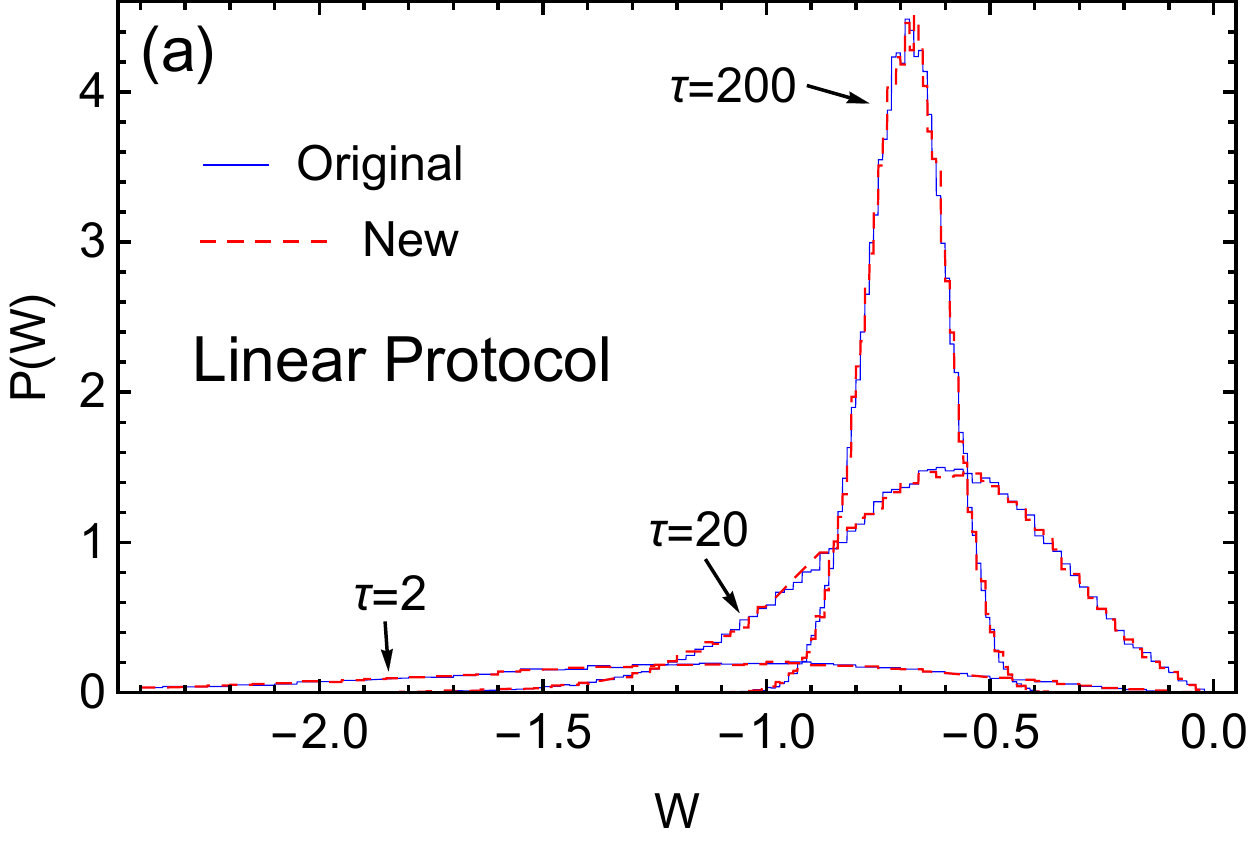}\\
        \includegraphics[width=7cm,clip]{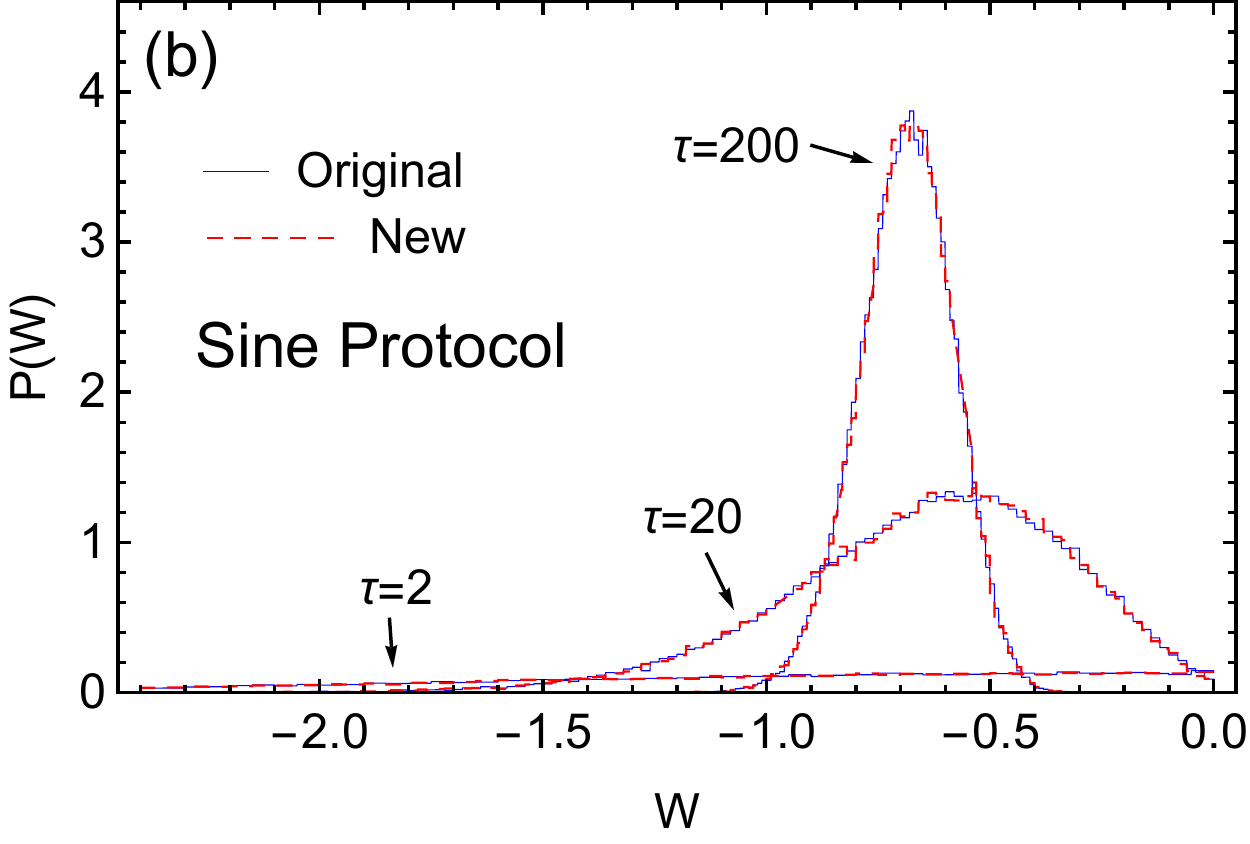}
   \end{center}
   \caption{Comparison of the work distributions obtained by the stochastic simulation based on the original EOM (Eq.~(1) in the main text) and work functional (Eq.~(2) in the main text) (blue solid line) and the new ones (Eq.~(\ref{EOM}) and Eq.~(\ref{functional})) (red dashed line), respectively for (a) the linear and (b) the sine protocols.}
\label{fig2}
\end{figure}

Besides the verification of the Jarzynski euqality based on the new EOM (\ref{EOM}) and work functional (\ref{functional}), which has been done in the main text, independent tests are required to help us completely confirm the validity of the subtle coordinate transformation. To this end, we directly compare the work distributions for the original as well as the new EOMs and work functionals. Moreover, since the linear protocol $\lambda_t=\lambda_0(1+t/\tau)$ is somehow a very special one with vanishing second order derivatives, we also perform this test for an additional nonlinear protocol $\lambda_t=2\lambda_0\sin\frac{\pi}{3}(\frac{t}{\tau}+\frac{1}{2})$. In particular, we numerically calculate the work distributions for three different time scales $\tau=2,20,200$, and all the results are presented in Fig.~\ref{fig2}. It is obvious that the work distributions obtained by the two different methods before and after the coordinate transformation agree perfectly with each other.

\section{An alternative derivation of the RFKE Eq.~(8) and the self-consistent check of its validity}
In the main text, we obtain the RFKE by first integrating the exact FKE with respect to $\xi$ then dropping two terms containing $h(\xi)$ due to the frequent collision approximation. In fact, we can derive the RFKE in a more intuitive way.

Considering a short time interval $dt$, in which $\lambda_t$ is almost unchanged but many collisions take place, we know that the number of the collisions at the right (movable) boundary can be estimated by $|\mathcal{P}_t|dt/2m\lambda_t$. When the system is well isolated, we roughly have
$\mathcal{\dot P}_t=-\dot\lambda_t\mathcal{P}_t/\lambda_t$, since $\mathcal{P}_t$ can be approximately regarded as $p_t$ with its sign modified, while each collision (at the right boundary) causes a momentum change with value $-2m\dot\lambda_t$. After including the dissipation effect, we can write down the following Langevin equation
\begin{equation}
\mathcal{\dot P}_t=-\left(\frac{\gamma}{m}+\frac{\dot\lambda_t}{\lambda_t}\right)\mathcal{P}_t+\sqrt{\frac{2\gamma}{\beta}}\eta_t,
\label{RLE}
\end{equation}
where a damping term and a stochastic term is added.

Also, noting that the work accumulated during a single collision is $-2\dot\lambda_t|\mathcal{P}_t|$, we immediately obtain the following reduced work functional
\begin{equation}
W_R[\mathcal{P}_t]=-\int^\tau_0dt\frac{\mathcal{P}^2_t\dot\lambda_t}{m\lambda_t}.
\label{RWF}
\end{equation}
 With a similar procedure to that used in deriving the exact FKE (first using the It\^o's lemma then modifying the generator), we finally obtain the RFKE (Eq.~(8) in the main text) from Eqs.~(\ref{RLE}) and (\ref{RWF}).

 The validity of the RFKE can be checked self-consistently. First, we could derive the JE by setting $s = \beta$. It can be checked that we have an exact solution $\varrho_\beta (\mathcal{P},t) = \varrho_\beta (\mathcal{P},0) \lambda_t / \lambda_0$ in this case, where $\varrho_\beta (\mathcal{P},0)$ is the probability density function of the normal distribution $N(0,m/\beta)$. After integrating $\varrho_\beta (\mathcal{P},t)$ over $\mathcal{P}$, we get $\psi_\beta (\tau) = \langle e^{- \beta W} \rangle = \lambda_\tau / \lambda_0$. Since the free energy of the piston system is $F(\lambda,\beta) = F(\lambda_0,\beta) - \beta^{-1} \ln (\lambda / \lambda_0)$, the validity of JE $\langle e^{-\beta W} \rangle = e^{- \beta \Delta F}$ is demonstrated.

Second, we can exactly solve the RFKE in the adiabatic limit $\gamma=0$. %For a Gaussian initial condition, the generating function turns out to be $\psi_s(t)=(1+\nu_t s/\beta)^{-1/2}, \nu_t=(\lambda_0/\lambda_t)^2-1$, implying the work distribution to be a Gamma distribution. This is consistent with the previous works based on totally different methods \cite{Crooks_2007,Gong_2014}.
 By writing ``adiabatic" here, we mean not only the absence of dissipation but also the quasistatic limit. Therefore, we can start from the RFKE with $\gamma=0$
\begin{equation}
\partial_t\varrho_s=\frac{\dot\lambda_t}{\lambda_t}\partial_{\mathcal{P}}\left(\mathcal{P}\varrho_s \right) + s\frac{\mathcal{P}^2\dot\lambda_t}{m\lambda_t}\varrho_s.
\end{equation}
The above partial differential equation is exactly solvable since  it is equivalent to
\begin{equation}
\partial_{\ln\lambda_t} \left(\varrho_s e^{\frac{s\mathcal{P}^2}{2m}-\ln\lambda_t}\right)=\partial_{\ln\mathcal{P}} \left(\varrho_s e^{\frac{s\mathcal{P}^2}{2m}-\ln\lambda_t}\right),
\end{equation}
which implies the following general solution
\begin{equation}
\varrho_s=\lambda_t e^{-\frac{s\mathcal{P}^2}{2m}}\varphi(\mathcal{P}\lambda_t),
\end{equation}
with $\varphi(\cdot)$ determined by the initial condition. In the quasistatic limit, the initial distribution function of $\mathcal{P}$ coincides with that of the original momentum $p$, i.e., a Gaussian distribution
\begin{equation}
\varrho_s|_{t=0}=\lambda_0e^{-\frac{s\mathcal{P}^2}{2m}}\varphi(\mathcal{P}\lambda_0)
%=\frac{1}{m\dot\lambda_0}[\mathrm{erf}(\sqrt{\frac{\beta}{m}}\mathcal{P}-\mathrm{erf}(\sqrt{\frac{\beta}{m}}\mathcal{P}-\sqrt{\beta m}\dot\lambda_0)]
=\sqrt{\frac{\beta}{2\pi m}}e^{-\frac{\beta\mathcal{P}^2}{2m}}.
\end{equation}
The above relation gives $\varphi(\cdot)$, and thus the special solution $\varrho_s$
\begin{equation}
\varrho_s=\frac{\lambda_t}{\lambda_0}\sqrt{\frac{\beta}{2\pi m}}e^{-\frac{\mathcal{P}^2}{2m}\left[(\frac{\lambda_t}{\lambda_0})^2(\beta-s)+s\right]}.
%=\mathrm{erf}(\sqrt{\frac{\beta}{m}}\frac{\mathcal{P}\lambda_t}{\lambda_0}-\sqrt{\beta m}\dot\lambda_0)]
\end{equation}

With the expression of $\varrho_s$ in hand, we can evaluate the generating function of work distribution by integrating out $\mathcal{P}$, obtaining
\begin{equation}
\psi_s(t)=\left(1+\frac{s}{\beta}\alpha_t\right)^{-\frac{1}{2}},
\end{equation}
where $\alpha_t=(\lambda_0/\lambda_t)^2-1$. Based on the above expression, by making use of the Laplace transformation formula $\mathscr{L}[u^{z-1}e^{-u}\theta(u)/\Gamma(z)]=(1+s)^{-z}$, with $\theta(u)$ and $\Gamma(z)$ respectively being the Heaviside step function and the Gamma function, we finally obtain the original work distribution function as follows
\begin{equation}
P(W)=\frac{\beta}{|\alpha_\tau|\Gamma(1/2)}\left(\frac{\beta W}{\alpha_\tau}\right)^{\frac{1}{2}-1}e^{-\frac{\beta W}{\alpha_\tau}}\theta\left(\frac{W}{\alpha_\tau}\right).
\end{equation}
This result is consistent with the previous work \cite{Crooks_2007}, where a totally different method (based on adiabatic invariance) is used. It is worth pointing out that in the so-called ``thermal wall model" \cite{Hatano_1998,Baule_2006,Engel_2013,Broeck_2015} one cannot reproduce the ``adiabatic" case by setting $\gamma=0$. Because in the ``thermal wall model" they did not use the standard EOMs of ST, i.e., Langevin equation and Fokker-Planck equation. We believe that our modeling of the isothermal piston is more consistent than the ``thermal wall model" in the framework of ST, and bridges the long-standing gap in the development of ST.

\end{document}